# The laws of physics do not prohibit counterfactual communication

Hatim Salih[1]✉, Will McCutcheon[2,3], Jonte R. Hance[2] and John Rarity[2]

It has been conjectured that counterfactual communication is impossible, even for post-selected quantum particles. We strongly challenge this by proposing precisely such a counterfactual scheme where—unambiguously—none of Alice's photons that correctly contribute to her information about Bob's message have been to Bob. We demonstrate counterfactuality experimentally by means of weak measurements, and conceptually using consistent histories—thus simultaneously satisfying both criteria without loopholes. Importantly, the fidelity of Alice learning Bob's bit can be made arbitrarily close to unity.

## INTRODUCTION

The prospect of communicating a message deterministically without exchanging physical carriers (e.g. photons) was first described by refs. [1,2], and demonstrated experimentally by refs. [3,4]. Apart from being extremely mind-boggling, this counterfactual communication raises deep questions about the nature of physical reality. For instance: "What carried the message across space?"; and, for the case of transmitted information itself being quantum, in the Salih14 protocol[5], "Did quantum bits vanish from one point in space only to discontinuously appear elsewhere?". No wonder, many prominent physicists were skeptical—most recently Griffiths, based on his Consistent Histories criterion for counterfactuality[6,7]; and Vaidman, based on his weak trace criterion for counterfactuality[8–10].

Around the same time the present paper was initially posted, Vaidman alongside colleague Aharonov posted an interesting paper conceding that counterfactual communication was possible—based on their modification of Salih et al.'s 2013 protocol to satisfy Vaidman's weak trace criterion for communication to be counterfactual[11], which is right. However, as we will show, their modification—unlike our scheme—does not by itself satisfy the Consistent Histories criterion for counterfactuality[6].

## RESULTS

### Setup

Our aim here is not to construct an efficient communication protocol, with each bit carried by a single photon, but rather to construct a communication protocol where:

i. Alice can determine Bob's bit choice with arbitrarily high accuracy, and;
ii. It can be shown unambiguously that Alice's post-selected photons have never been to Bob.

Also, our purpose here is not to construct a secure communication protocol—an eavesdropper may be able to exploit our protocol to obtain the information Bob sends without Alice or Bob realising.

Consider our proposed protocol setup in Fig. 1, which includes the equivalent of one outer cycle of Salih et al.'s (Michelson-type) protocol laid-out sequentially in time, as in ref. [12]. The two underlying principles here are interaction-free measurement[13] and the quantum Zeno effect[14]. Here's how the setup works. Alice sends a *H*-polarized photon from photon source **S**, whose polarization is then rotated by the action of polarization rotator HWP1, before polarizing beam splitter PBS passes the *H* part along arm **A**, while reflecting the *V* part along arm **D**. (All PBSs transmit *H* and reflect *V*.) The *V*-polarized component in **D** then encounters a series of polarization rotators HWP2, each affecting a small rotation, and polarizing beamsplitters PBS, whose collective action is to rotate polarization from *V* to *H*, if Bob does not block the transmission channel. In this case, this component is passed straight towards detector $D_3$. If the photon is not lost to $D_3$ then we know that it has traveled along arm **A** instead, in which case it passes through two consecutive PBS's on its way to $D_0$. What happens if Bob instead blocks the transmission channel? Provided that the photon is not lost to Bob's blocking devices, the part of the photon superposition that was in arm **D** at time $t_1$ is now, after last HWP2, in arm **D**, *V*-polarized. It is then reflected by two consecutive PBS's on its way towards detector $D_1$. A click at detector $D_1$ corresponds uniquely to Bob blocking the channel. But there's a chance that the photon component that has traveled along **A** causes detector $D_0$ to incorrectly click. For example, given that Alice had initially rotated her photon's polarization after time $t_0$ such that it is in arm **A** with probability 1/3, and in arm **D** with probability 2/3, and given a large number of HWP2's such that the chance of losing the photon to Bob's blocking device is negligible, then it is straight forward to calculate that the accuracy of detector $D_0$ is 75%, in contrast to 100% accuracy for $D_1$, with half the photons being lost on average. Importantly, accuracy can be made arbitrarily close to 100%, by HWP1 initially rotating the photon's polarisation closer to *V*, at the expense of more photons being lost.

Note, the postselection in the protocol is passive—it happens without any communication with Bob. The post-selection process simply corresponds to the instances of photons arriving at Alice's detectors $D_0$ or $D_1$. The protocol works because when Bob blocks there is a much higher probability for a photon to arrive at detector $D_1$ than detector $D_0$. When Bob doesn't block, and ignoring device imperfections, all post-selected detections happen at detector $D_0$, with the photon having arrived directly from Alice. Detector $D_3$ is not needed for the communication, and is simply a loss channel. In both cases where Alice obtains information, the photon stays in her lab, and thus provably never goes to Bob.

[1]Quantum Technology Enterprise Centre, HH Wills Physics Laboratory, University of Bristol, Tyndall Avenue, Bristol BS8 1TL, UK. [2]Quantum Engineering Technology Laboratory, Department of Electrical and Electronic Engineering, University of Bristol, Woodland Road, Bristol BS8 1UB, UK. [3]Institute of Photonics and Quantum Science, School of Engineering and Physical Sciences, Heriot-Watt University, Edinburgh EH14 4AS, UK. ✉email: salih.hatim@gmail.com

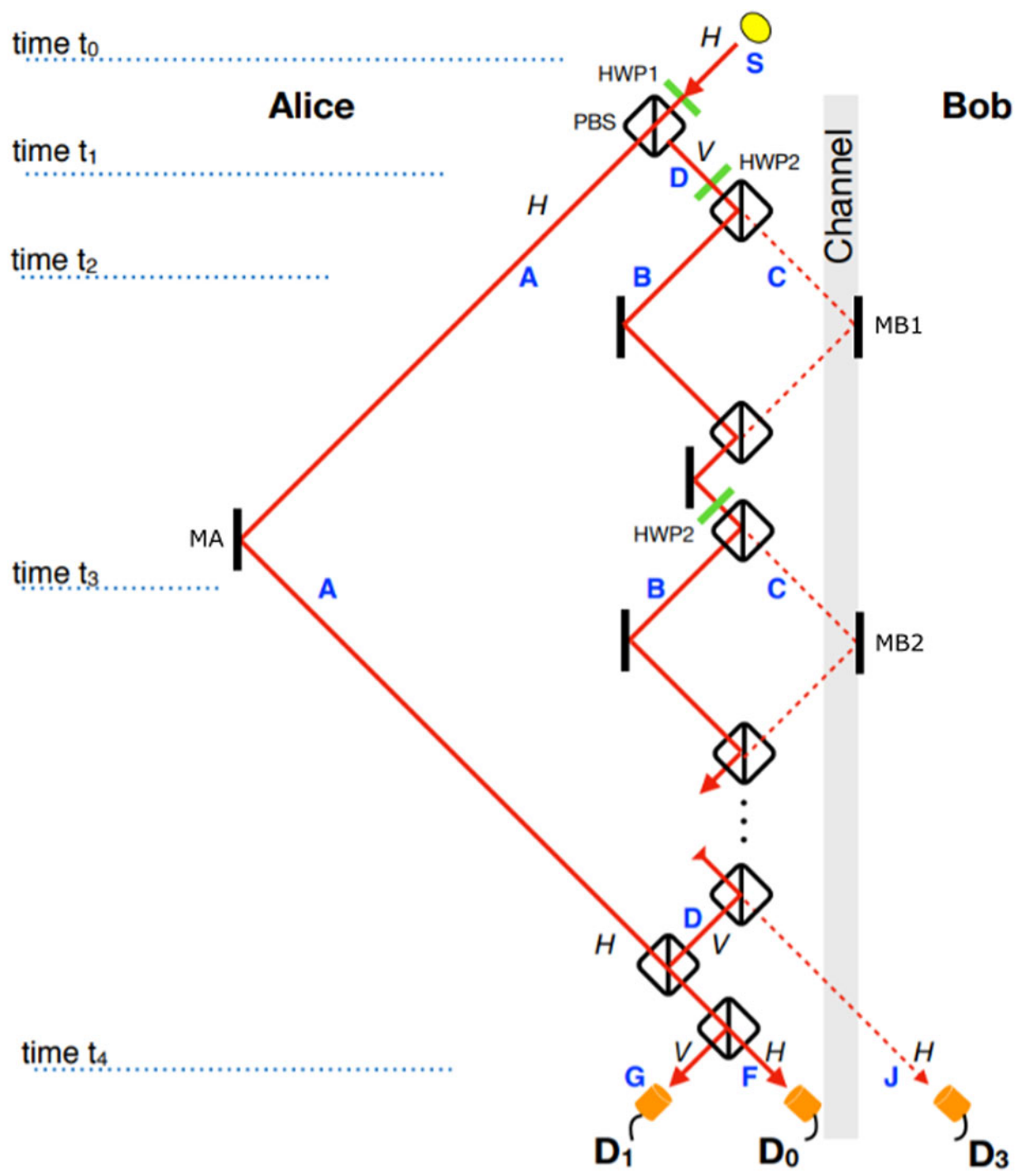

**Fig. 1 Schematic setup for our one cycle implementation of Salih et al.'s 2013 counterfactual communication protocol.** See ref. [1]. All beamsplitters are polarising beamsplitters (PBSs), transmitting $H$-polarised and reflecting $V$-polarised light. We want to know if photons detected at Alice's $D_0$ have been to Bob on the right-hand side. We place detector $D_0$ at the bottom, rather than immediately after the topmost PBS, so that the setup exactly includes the equivalent of one cycle of Salih et al.'s Michelson-type protocol laid-out sequentially in time[12]. This allows the conclusions drawn for this one cycle to be applicable to any of the concatenated cycles from the 2013 counterfactual communication protocol.

## Demonstrating counterfactuality

Now we turn to the question of whether Alice's post-selected photon has ever been to Bob. It is accepted that for the case of Bob blocking the transmission channel, Alice's photon could not have been to Bob—otherwise the photon would have been absorbed by Bob's blockers (which act as loss modes from path **C** in this protocol). It is the case of Bob not blocking the channel that is interesting.

## Consistent Histories

We first consider the question from a consistent histories (CH) viewpoint, building on the analysis in ref. [12]. While we do not unreservedly advocate the Consistent Histories interpretation, we still engage with it here within the frame of Consistent Histories, as it has been used by Griffiths to question the counterfactuality of counterfactual communication protocols[7]. See ref. [7] for a thorough explanation of CH. By constructing a family $\mathcal{Y}$ of consistent histories (which we will shortly explain the meaning of) between an initial state and a final state, that includes histories where the photon takes path **C**, we can ask what the probability of the photon having been to Bob is. Our setup allows us to do just that.

$$\mathcal{Y}: S_0 \otimes H_0 \odot \{A_1 \otimes I_1, D_1 \otimes I_1\} \odot$$
$$\{A_2 \otimes I_2, B_2 \otimes I_2, C_2 \otimes I_2,\} \odot$$
$$\{A_3 \otimes I_3, B_3 \otimes I_3, C_3 \otimes I_3,\} \odot F_4 \otimes H_4$$

where $S_0$ and $H_0$ are the projectors onto arm **S** and polarization $H$, respectively, at time $t_0$. $A_1$ and $I_1$ are the projectors onto arm **A** and the identity polarization $I$ at time $t_1$, etc. The curly brackets contain different possible projectors at that given time. A history then consists of a sequence of projectors, at successive times. This family of histories therefore consists of a total of 18 histories. For example, the history $(S_0 \otimes H_0) \odot (A_1 \otimes I_0) \odot (A_2 \otimes I_2) \odot (A_3 \otimes I_3) \odot (F_4 \otimes H_4)$ has the photon traveling along arm **A** on its way to detector $D_0$. Each history has an associated chain ket, whose inner product with itself gives the probability of the sequence of events described by that particular history. Here's the chain ket associated with the history we just stated, $|S_0 \otimes H_0, A_1 \otimes I_1, A_2 \otimes I_2, A_3 \otimes I_3, F_4 \otimes H_4\rangle = (F_4 \otimes H_4)T_{4,3}(A_3 \otimes I_3)T_{3,2}(A_2 \otimes I_2)T_{2,1}(A_1 \otimes I_1)T_{1,0}|S_0H_0\rangle$, where $T_{1,0}$ is the unitary transformation between times $t_0$ and $t_1$, etc. By applying these unitary transformations and projections, we see that this chain ket is equal to, up to a normalization factor, $|F_4H_4\rangle$.

A family of histories is said to be consistent if all its associated chain kets are mutually orthogonal. It is straight forward to verify that for the family $\mathcal{Y}$ above, each of the other 17 chain kets is zero. For example, the chain ket $|S_0 \otimes H_0, D_1 \otimes I_1, C_2 \otimes I_2, I_3 \otimes I_3, F_4 \otimes H_4\rangle = (F_4 \otimes H_4)T_{4,3}(I_3 \otimes I_3)T_{3,2}(C_2 \otimes I_2)T_{2,1}(D_1 \otimes I_1)T_{1,0}|S_0H_0\rangle = (F_4 \otimes H_4)T_{4,3}(I_3 \otimes I_3)T_{3,2}(C_2 \otimes I_2)T_{2,1}|D_1V_1\rangle = (F_4 \otimes H_4)T_{4,3}(I_3 \otimes I_3)T_{3,2}|C_2H_2\rangle = (F_4 \otimes H_4)T_{4,3}(|C_3H_3\rangle + |B_3V_3\rangle) = (F_4 \otimes H_4)(|G_4V_4\rangle + |J_4H_4\rangle)$, up to a normalization factor. Because projectors $F_4$, $G_4$, and $J_4$ are mutually orthogonal, this chain ket is zero.

Family $\mathcal{Y}$ is therefore consistent.

This means, using $\mathcal{Y}$, we can ask the question of whether the photon has been to Bob. CH gives a clear answer: Since every history in this family, except the one where the photon travels along arm **A**, is zero, we can conclude that the photon has never been to Bob.

(Note that, when considering the time evolution of histories ending up at detector $D_1$, we can get a non-zero ket for ones where the photon travels to Bob, e.g. a history where the photon is on path **C** at time $t_2$. However, for the case in question where Bob doesn't block, we know that except for experimental imperfections, only detectors $D_0$ and $D_3$ can click—in other words detector $D_1$ never clicks. This is because the coherent evolution in the inner interferometer-chain ensures that any photon exiting the chain is $H$-polarised, and as such cannot go to $D_1$. We therefore apply postselection to "manually" exclude all unphysical histories ending at $D_1$. We caution that this is only possible because this detection forms a final measurement).

It can straightforwardly be seen that any history containing a projector $C_i$, in any family of histories where the photon ends up in arm **F** (regardless of coarse- or fine-graining), has zero probability due to the final state projection $F_4 \otimes H_4$. Therefore, for all relevant consistent families, we can conclude the photon has never been to Bob.

## Weak trace

We now ask the same question in the weak trace language. Weak measurements[15], as the name suggests, consist of making measurements so weak that their effect on individual particles is smaller than uncertainty associated with the measured observable, and is therefore indistinguishable. These weak measurements on pre- and post-selected states, have related, usually well-defined quantities called weak values, for which only particles that start in a particular initial state and are found in a particular final state are considered. By looking at a large-enough number of such particles, these measurements result in definite, predictable pointer-shifts in the measuring device, corresponding to the weak values.

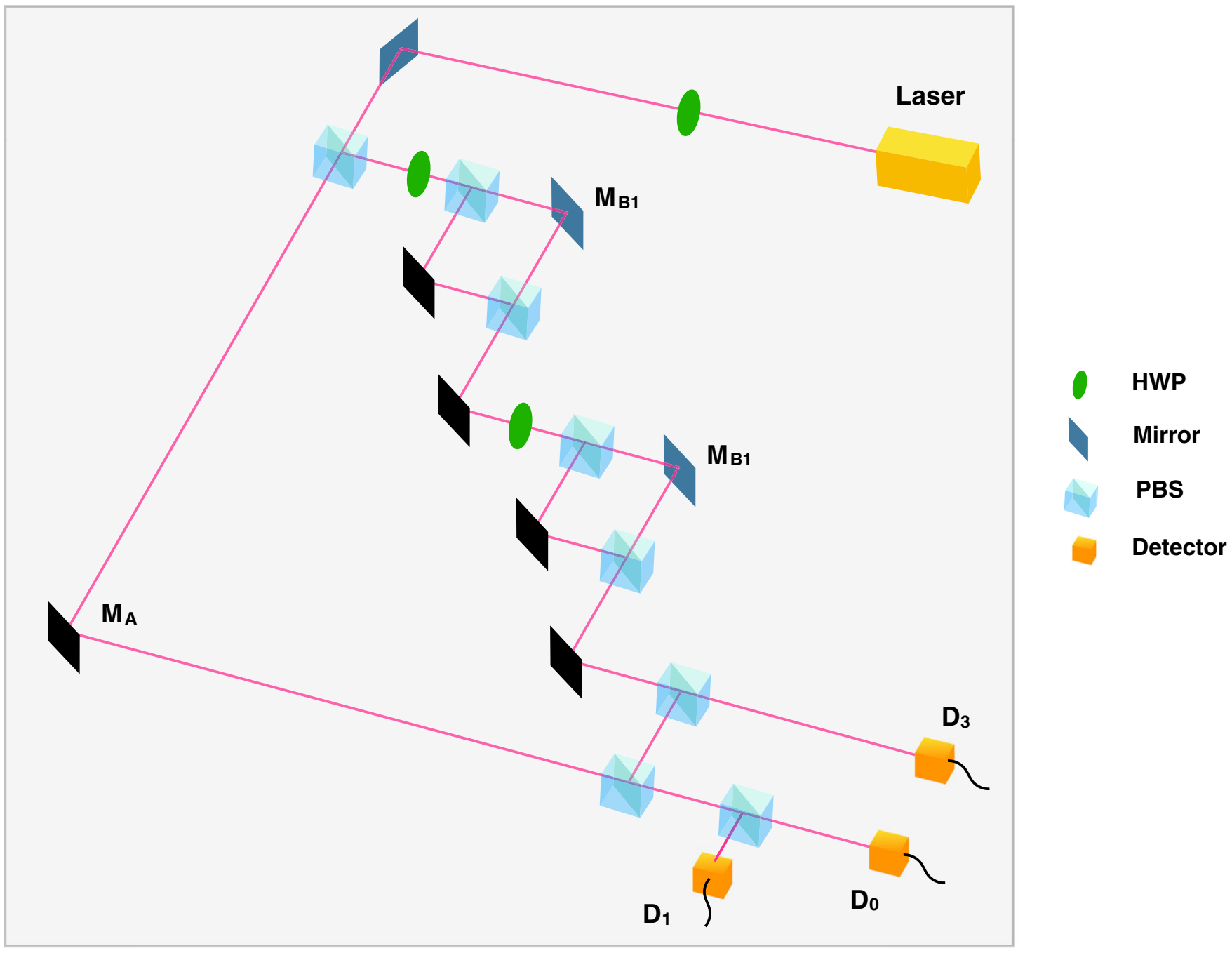

**Fig. 2 3D depiction of experimental setup.** This is based on the setup in Fig. 2. $M_A$, $M_{B1}$ and $M_{B2}$ are MEMS mirrors oscillating at different frequencies. If a frequency associated with a given mirror is absent from the power spectrum at detector $D_0$, then according to Vaidman's weak trace approach, we know that photons detected at $D_0$ have not been near that mirror.

An elegant way of predicting non-zero weak values of the position operator, at least as a first order approximation, is the two state vector formulation, TSVF[16]. If the initial state evolving forward in time overlaps at a given point with the final state evolving backward in time, then the weak value of the particle number at that point is nonzero. Vaidman associates these non-zero weak values with a weak trace, claiming a quantum particle has been wherever there is this weak trace. While we do not unreservedly advocate this point of view (which has been extensively analysed[17–20]), we still engage with it here within the frame of weak measurement, as it has been used to challenge the counterfactuality of counterfactual communication protocols[8,9,21].

Let's apply this to our setup. The pre-selected state is that of the photon in arm **S**, *H*-polarized. And the post-selected state, for the case in question of Bob not blocking, is that of the photon in **F**, also *H*-polarized. Consider weak measurements where Bob's mirrors, $M_{B1}$ and $M_{B2}$, are made to vibrate at specific frequencies, before checking if these frequencies show up at a detector $D_0$ capable of such measurement[18,22,23]. The forward evolving state from **S** is clearly present at Bob's, because of the photon component directed by the action of HWP1 and PBS along arm **D**. What about the backward evolving state from **F**? A *H*-polarized photon traveling from **F** will pass through the two consecutive PBS's along arm **A**, away from Bob. Since, the forward evolving state and the backward evolving state do not overlap at Bob, a weak measurement, at least as a first order approximation, will be zero.

We now show that any weak measurement at Bob will be zero —not just to a first order approximation. Consider a weak measurement where Bob vibrates one or more of his mirrors (as in ref. [22]). This disturbance will cause the part of the photon superposition in arm **D**, after the last HWP2, which can only be *V* polarized, to be nonzero. This small *V* component will be reflected by two PBS's towards $D_1$, and crucially, away from detector $D_0$. Bob's action has no way of reaching Alice's post-selected state: The photon has never been to Bob.

We performed this experiment using a version of the setup in Fig. 1, which we show in Fig. 2, with two inner M-Z interferometers within one outer cycle of Salih et al's protocol. The polarising beamsplitters used are ThorLabs PBS251, and the half-wave plates used are ThorLabs AHWP05M-600 - all three half wave plates are tuned with their fast axis at an angle to the normal such that the polarisation of the light is rotated by $\pi/2$ (i.e. $H \rightarrow (H+V)/\sqrt{2}$). In this setup, the single photon source is replaced by a continuous-wave diode laser [635nm ThorLabs LDM635], and the three vibrating mirrors are micro-electro-mechanical systems (MEMS) mirrors [Hamamatsu S12237-03P] weakly oscillating sinusoidally in the horizontal plane—Alice's mirror ($M_A$) at 29Hz, Bob's first mirror ($M_{B1}$) at 13Hz, and his second mirror ($M_{B2}$) at 19Hz (chosen so not to be harmonics of each other). This oscillation is made sufficiently weak so as not to disturb the counterfactual properties of the system. This was done by having maximal 0.01 mm movement detected over a 5 mm beam diameter at the detectors. The other mirrors in the set-up are all standard ThorLabs MRA25-E02 mirrors. The detectors used for $D_0$, $D_1$ and $D_3$ are segmented quadrant position-sensing photodetectors [ThorLabs PDQ80A], sampled by a LeCroy Wavesurfer 452 at a rate of 25 KHz for 5 s. We applied a Fast Fourier transform to the position signal as a function of time, to observe the spectrum of oscillation frequencies at $D_0$, $D_1$ and $D_3$. As can be seen from Fig. 3a, there is an oscillation at $D_0$ from Alice's mirror but not from either of Bob's mirrors. This shows that the weak measurement at Bob is zero, thus demonstrating experimentally the counterfactuality of the protocol.

Note, in Fig. 3c, the peak for $M_{B1}$ and its harmonic at $D_3$ are larger than that for $M_{B2}$, as $M_{B1}$ is further away from the detectors than $M_{B2}$. Note also that the peak for $M_A$ in $D_1$, shown in Fig. 3b, is close to the noise level as it is an erroneous signal, caused by light from Alice's path leaking into $D_1$. However, the fact we can see this error signal at $D_1$ in Fig. 3b, but not the error signals at $D_1$ from $M_{B1}$ and $M_{B2}$, shows that in all cases only a negligible amount of light (i.e. lower than noise) leaks from Bob back to Alice.

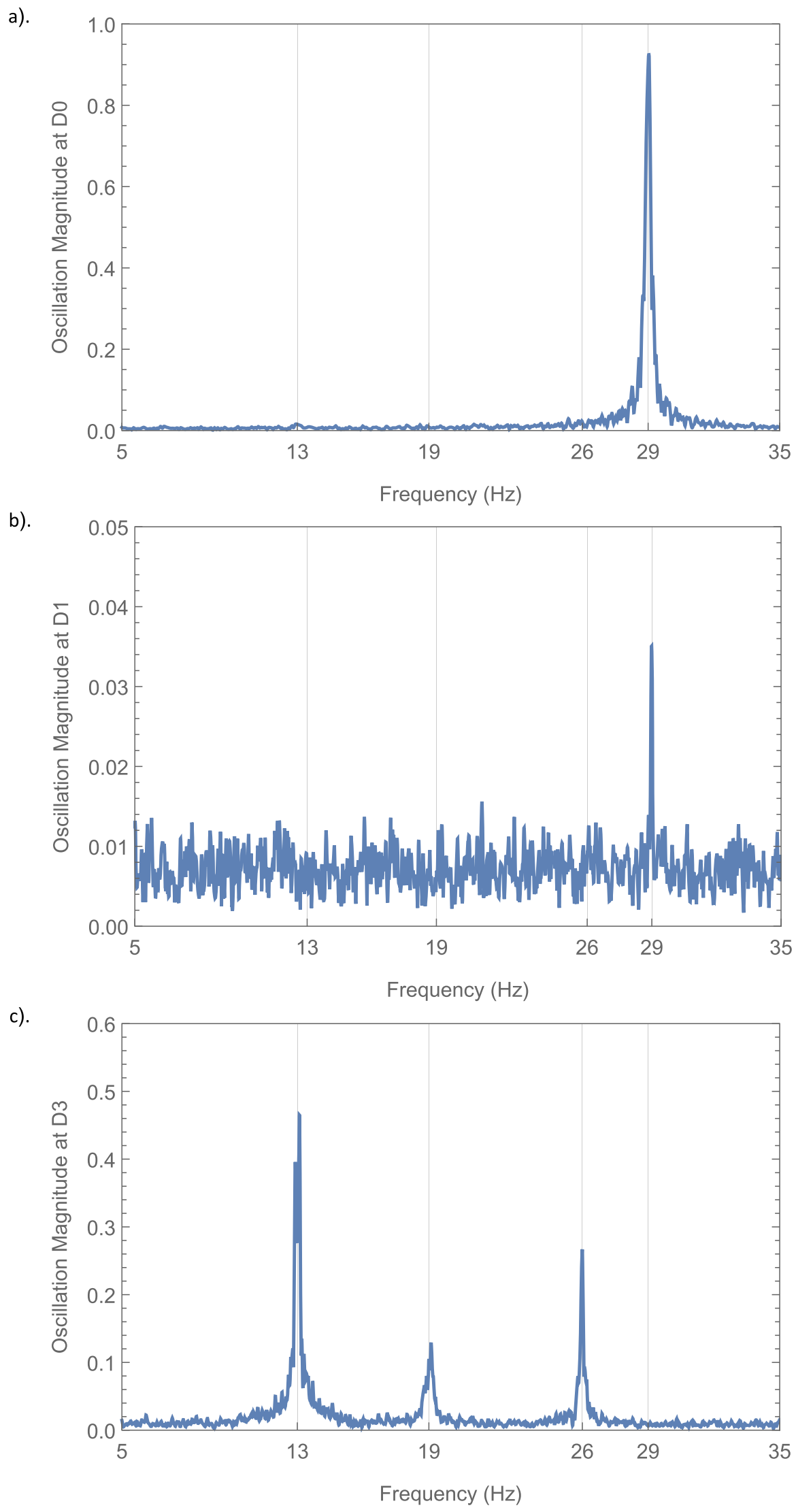

**Fig. 3 Weak measurement tagging showing no weak trace from Bob's mirrors at detector $D_0$ or $D_1$.** Fourier transform of position with respect to time, of light beam incident on detectors $D_0$ (**a**), $D_1$ (**b**) and $D_3$ (**c**). $D_0$ and $D_1$ show the oscillation from Alice's mirror (at 29Hz), but unlike $D_3$ do not show the oscillations from Bob's two mirrors (at 13Hz and 19Hz, with a second harmonic at 26Hz), proving via weak measurement that no light that goes to Bob's mirrors ends up at either $D_0$ or $D_1$.

## DISCUSSION

Having shown that any photon detected by Alice at $D_0$ or $D_1$ will have never been to Bob, we now look in depth at the probabilities of success, and Alice receiving the correct bit-value. In each round of the proposed experiment Bob chooses a bit, $X$, he would like to communicate to Alice. He blocks (does not block) his channel when $X = 0$ ($X = 1$). Alice then prepares a single photon, passes it through the system and it is either detected in one of the detectors $D_0$, $D_1$ and $D_3$, or is lost to Bob's blocking device. If Alice detects the photon in either $D_0$ or $D_1$, then the round was successful and Alice assigns the estimated values $X_{est} = 0$ and $X_{est} = 1$ to detections in $D_0$ and $D_1$ respectively. If the round was not successful another round is performed until she obtains a successful outcome. The post-selected data she obtains displays clear communication from Bob despite the fact that, as we have shown, the postselected photons never passed through the communication channel to Bob. Furthermore, by tuning the initial half-wave plate, the system can be tuned to achieve a postselected success probability arbitrarily approaching unity, at the expense of decreasing the post-selection probability.

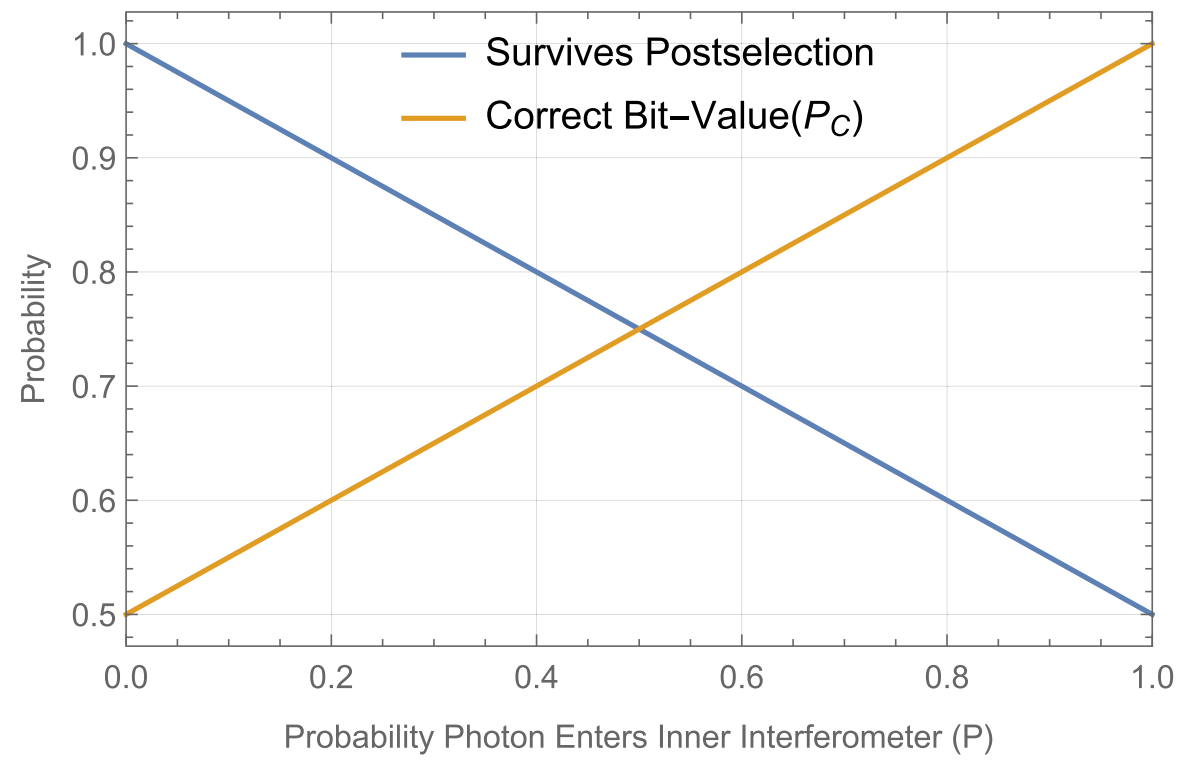

**Fig. 4 Postselection success and accuracy probabilities.** The total postselection survival probability (blue), and probability of post-selected correct outcome, $P_C$ (orange), for a given bit sent from Bob to Alice in the infinite inner-cycles case of the protocol, plotted against $P$, the probability of the photon entering the inner interferometer chain.

We explore the success of the scheme in terms of the free parameter $P$, the raw probability the photon would be found in the right half of the setup, determined by the setting of HWP1.

The raw conditional probabilities of detection in each of the detectors given Bob blocking and not blocking his side of the channel, for the infinite inner cycle version of the protocol, are: for blocking, a probability of detection in $D_0$ of $1 - P$, in $D_1$ of $P$, and in $D_3$ (lost) of 0; for not blocking, a probability of detection in $D_0$ of $1 - P$, in $D_1$ of 0, and in $D_3$ (lost) of $P$.

Note that the protocol has an error chance that varies depending on whether Bob sends a 0-bit or a 1-bit. This, however, is only a feature of the one-outer-cycle form of counterfactual communication we give here when two or more outer cycles are used instead, the error probability in the protocol is 0, and therefore doesn't depend on the bit sent (see the Appendix of ref. [24] for more discussion on unequal bit error rates in counterfactual communication).

Consider for now the limit in which the probability of losing the photon to Bob's blocking apparatus vanishes. Since post-selection for the case of not blocking only succeeds with probability $P(D_0|NB) = (1 - P)$, Alice must perform the experiment many times to get a successfully postselected event. This achieves communication in the postselected data since the conditional probability of blocking given detection events at $D_0$ decreases with $P$ increasing.

We assume that on average Bob encodes as many zeros as he does ones. From the raw detection probabilities, the probability of the protocol giving an outcome, that is the post-selection succeeding, is $(2 - p)/2$.

We then find the probabilities of the postselected detection events: the probability for detecting in $D_0$ when blocking is $(1 - P)$, and when not blocking is 1; the probability for $D_1$ when blocking is $P$, and when not blocking is 0. Therefore after post-selection, the probability of correct outcome of $P_c = (1 + P)/2$.

We see that in the limit $P \rightarrow 1$ the protocol becomes deterministic, however the probability of postselection vanishes.

In Fig. 4 we plot the overall probability of successful postselected outcome and postselection probability, for different values of the probability $P$ of the photon entering the inner interferometer chain. Notably, for $P = 1/2$ postselection succeeds (photon arrives at Alice) with 3/4 probability, and, if it arrives, is correct (is the bit Bob sent) with 3/4 probability. Increasing $P$ to 2/3, the likelihood of successful postselection drops to 2/3 whilst the probability of being correct increases to 5/6.

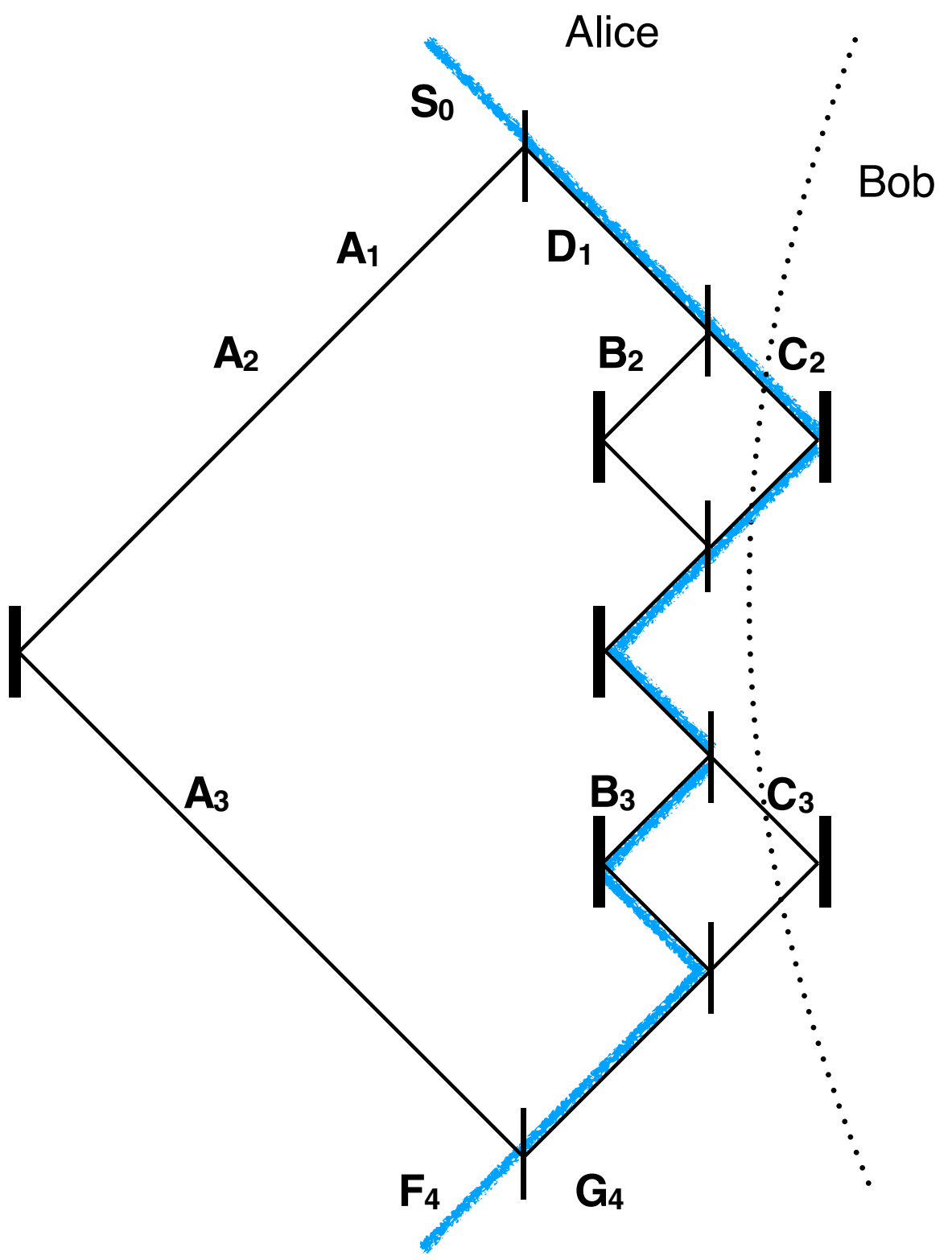

**Fig. 5 Aharonov and Vaidman's protocol overlaid with a counterfactuality-violating history.** The modified protocol proposed by Aharonov and Vaidman in ref. [11], of which we show one cycle, cannot be said to be counterfactual from a consistent histories viewpoint. This can be seen from the series of projections, or history, highlighted in blue. The thin vertical lines represent non-polarising beamsplitters, thick vertical lines represent mirrors, and the blue path shows an example of a history where, when Bob doesn't block, the photon travels to Bob and back to Alice's relevant detector with nonzero probability. See text for mathematical details.

Finally, let's illustrate our findings using an amusing scenario. Imagine an outcome-obsessed lab director in charge of this experiment, who is quite happy firing Alice and Bob if a single run of the experiment fails, replacing them with a fresh pair of experimentalists, to start all over, also nicknamed Alice and Bob. The task for any Alice and Bob pair is to communicate a 16-bit message, one bit at a time. Assume the experiment is set up such that the chance of any given run failing is 1/4. Therefore, in order to successfully communicate the 16-bit message, the lab director has to, on average, go through around 100 pairs of experimentalists—which the director secretly enjoys. Each new pair of experimentalists is provided with a new message. Eventually, a lucky Alice and Bob manage to communicate their message (bit accuracy will be 75% on average.) Now the question for the successful pair is: Has any of Alice's photons been to Bob while communicating the message? The answer, as we have shown, is an emphatic no.

Aharonov and Vaidman posted an interesting paper, just before we first posted the present paper on the arXiv, suggesting a way to modify a version of Salih et al.'s 2013 protocol that does not use polarisation. Their modification has the effect of satisfying Vaidman's weak trace criterion for counterfactuality for both bits. The scheme has the advantage over ours of reducing the chance of a communication error caused by imperfect interference in the inner interferometers when Bob encodes bit 0, thus satisfying the weak trace criterion even for erroneous bit-clicks, which we are not concerned about. However, unlike the scheme we give above, their modification fails the Consistent Histories criterion for counterfactuality. In Fig. 5 we give an example of a history where the photon goes to Bob, where the associated chain-ket is nonzero. More precisely, take the histories family,

$$\begin{aligned}\mathcal{Y}':&S_0 \odot \{A_1, D_1\} \odot \\ &\{A_2, B_2, C_2,\} \odot \\ &\{A_3, B_3, C_3,\} \odot F_4\end{aligned} \tag{1}$$

The chain ket associated with the path highlighted in blue in Fig. 5 is

$$\begin{aligned}|S_0, D_1, C_2, B_3, F_4\rangle &= (F_4)T_{4,3}(B_3)T_{3,2}(C_2)T_{2,1}(D_1)T_{1,0}|\mathrm{S}_0\rangle = \\ &(F_4)T_{4,3}(B_3)T_{3,2}(C_2)T_{2,1}|\mathrm{D}_1\rangle = (F_4)T_{4,3}(B_3)T_{3,2}|\mathrm{C}_2\rangle = \\ &(F_4)T_{4,3}|\mathrm{B}_3\rangle = |\mathrm{F}_4\rangle\end{aligned} \tag{2}$$

up to a nonzero normalization factor. Similarly, the chain ket associated with path $A$

$$\begin{aligned}|S_0, A_1, A_2, A_3, F_4\rangle &= (F_4)T_{4,3}(A_3)T_{3,2}(A_2)T_{2,1}(A_1)T_{1,0}|\mathrm{S}_0\rangle = \\ &(F_4)T_{4,3}(A_3)T_{3,2}(A_2)T_{2,1}|\mathrm{A}_1\rangle = (F_4)T_{4,3}(A_3)T_{3,2}|\mathrm{A}_2\rangle = \\ &(F_4)T_{4,3}|\mathrm{A}_3\rangle = |\mathrm{F}_4\rangle\end{aligned} \tag{3}$$

is also nonzero up to a nonzero normalization factor. The two chain kets are not orthogonal, which means that the family of histories is not consistent. The original Aharonov–Vaidman setup therefore cannot be said to be counterfactual from a consistent histories point of view. Nonetheless, we agree that Aharonov and Vaidman's modification presents important progress, as it robustly passes the weak trace criterion for counterfactuality (as shown in ref. [25]). It has thus been included as an additional element by some of the present authors in devices proposed for counterfactual communication (e.g. refs. [19,26–28]).

Likewise, Aharonov-Vaidman's modification can be incorporated straightforwardly in our present setup. The way to do this is by simply repeating the inner-cycles sequence twice, between the arms marked **D**. This would eliminate the weak trace even for erroneous detector-clicks, while still passing the consistent histories test.

In summary, we have shown both theoretically and experimentally that, given post-selection, sending a message without exchanging any physical particles is allowed by the laws of physics. What carries this information, however, remains a hot topic of research[19,29,30].

## DATA AVAILABILITY

Data obtained during experiments undertaken during this experiment is available at https://doi.org/10.5281/zenodo.5666675.

## CODE AVAILABILITY

The code generated to analyse the protocol is available at https://doi.org/10.5281/zenodo.5666675.

## ACKNOWLEDGEMENTS

H.S. thanks Li Yu-Huai for useful feedback on the basic concept, received in autumn 2016. The authors thank Paul Skrzypczyk for useful discussions on the issue, David Lowndes for experimental components and advice, Gary F Sinclair for help with fast Fourier transform code, and Huxley Sessa for help with experimental prototyping. This work was supported by the Engineering and Physical Sciences Research Council (Grants EP/P510269/1, EP/R513386/1, EP/L024020/1 and the UK Quantum Communications Hub funded by EP/M013472/1 and EP/T001011/1).

## AUTHOR CONTRIBUTIONS

(CRediT)— H.S.: conceptualisation, investigation, formal analysis, methodology, visualisation, writing (original draft), writing (review and editing), supervision, project administration; W.M. and J.R.H.: investigation including experimental, formal analysis, methodology, software, visualisation, writing (original draft), writing (review and editing), project administration; J.R.: supervision, writing (review and editing), funding acquisition, resources.

## COMPETING INTERESTS

The authors have no competing interests.

## ADDITIONAL INFORMATION

**Correspondence** and requests for materials should be addressed to Hatim Salih.

**Reprints and permission information** is available at http://www.nature.com/reprints

**Publisher's note** Springer Nature remains neutral with regard to jurisdictional claims in published maps and institutional affiliations.